\documentclass{elsart}
\usepackage {graphicx,amssymb}
\begin{document}
\begin{frontmatter}
\title{Interference between Coulomb and hadronic scattering in elastic
high-energy nucleon collisions}
\author{Vojt\v{e}ch Kundr\'{a}t, 
Milo\v{s} Lokaj\'{\i}\v{c}ek}
\footnote{e-mail: kundrat@fzu.cz,lokaj@fzu.cz}
\address{Institute of Physics,
AS~CR, 182 21 Prague 8, Czech Republic}
\maketitle
\begin{abstract}
The different models of elastic nucleon scattering
amplitude will be discussed. Especially, the preference of the
more general approach based on eikonal model will
be summarized in comparison with the West and Yennie
amplitude that played an important role in analyzing
corresponding experimental data in the past.  

\end{abstract}  
\end{frontmatter}
\section{Introduction}
\label{sec1}

Elastic nucleon scattering at high energies 
represents probably the most extensive and
the most precise ensemble of available 
experimental data enabling to perform
very accurate analysis in a broad region 
of the four momentum transfer squared $t$. 
The elastic scattering of nucleons is 
realized mainly due to the strong hadronic 
interactions. However, in the case of charged 
hadrons the elastic scattering is also realized 
due to the Coulomb interactions, which play
an important role mainly at small $|t|$. 

The influence of both the strong and electromagnetic 
interactions in the elastic scattering of protons 
by a nucleus was first investigated by Bethe \cite{beth}. 
Using the semi classical WKB approximation he derived 
that the total elastic amplitude $F^{C+N}(s,t)$ 
can be written in principle as the sum of hadronic 
amplitude $F^{N}(s,t)$ and of Coulomb amplitude 
$F^{C}(s,t)$ known from QED which were mutually 
bound by a relative phase; here $s$ is the square 
of the CMS energy. Higher corrections to the relative
phase were then obtained by Islam \cite{isl1} by 
improving the approximations.

The first generalization of the relative phase
within the relativistic theory was given by 
Soloviev \cite{solo} using the methods of QED
and also by Rix and Thaler \cite{rix}. Applying 
the methods of Feynman diagram technique 
(one-photon exchange) Locher \cite{loch} 
and mainly West and Yennie \cite{west}
derived a more general expression for 
the relative phase in the case of 
charged point-like particles as 
\begin{equation}
\Phi (s,t) = \mp \bigg [ \ln \bigg ( {{-t} \over s} \bigg ) -
\int_{-4 p^2}^{0} {{ d t} \over {|t - t'|}} \bigg ( 
1 - {{F^N(s,t')} \over {F^N(s,t)}} \bigg ) \bigg ],
\label{wy1}
\end{equation}
where $p$ is the value of the momentum in the CMS.

Similar expression as (\ref{wy1}) was derived by
Franco \cite{fra1} within the framework of eikonal 
model and by Lapidus and co-workers \cite{lapi}. 
Franco and Varma \cite{fra2} generalized this method 
also for the hadron-nucleus and nucleus-nucleus
scattering. The eikonal model approach was used 
later by Cahn \cite{cahn} who took into account also 
the form factors in order to describe the electromagnetic
nucleon collisions as the extended objects with the accuracy 
up to the terms linear in $\alpha$. Higher order corrections 
to the relative phase were then calculated by Selyugin 
\cite{sely} and by Kopeliovich and Tarasov \cite{kope}. 
The influence of spins was studied within the impact
parameter formalism applying the helicity amplitudes 
by Buttimore, Gotsman and Leader \cite{butt}.
All these authors obtained the explicit analytical 
formulas for the relative phase describing 
the elastic scattering at small values of $|t|$ by
introducing some further assumptions specifying the $t$ 
dependence of the hadronic amplitude $F^{N}(s,t)$. 
The impact parameter representation of the scattering 
amplitude used by the mentioned authors has been valid 
at very high energies and small $|t|$ only. 

In all these approaches the total elastic scattering
amplitude could be at the end written (after introducing
some other approximations) in a very simplified 
form proposed already in Ref. \cite{west}. After 
adding the two form factors $f_1(t)$ and $f_2(t)$ 
corresponding to individual colliding nucleons
to the Coulomb scattering amplitude the total 
elastic scattering amplitude has been written as
\begin{eqnarray}
F^{C+N}(s,t) &=&
F^{C}(s,t) e^{i\alpha \Phi (s,t)} + F^{N}(s,t)  =
\nonumber  \\ 
&=& \pm {\alpha s \over t} f_1(t)f_2(t)e^{i\alpha \Psi}+
{\sigma_{tot} \over {4\pi}} p\sqrt {s} (\rho+i)e^{Bt/2},
\label{wy2}
\end{eqnarray}
where $\alpha=1/137.036$ is the fine structure constant. 
The upper (lower) sign corresponds to the scattering of 
particles with the same (opposite) charges. The quantity 
$\rho$ and the diffractive slope $B$ in the second term 
are independent of $t$; together with the total cross 
section $\sigma_{tot}$ they can be energy dependent only 
and characterize the elastic hadron scattering at given 
energy.

It has been shown then by Locher \cite{loch} 
and West and Yennie \cite{west} that the relative
phase $\alpha \Phi(s,t)$ is to depend on the hadronic
amplitude in a rather simple way; they have obtained
\begin{equation}
\Phi (s,t) = \mp \bigg [ \ln \bigg ({{-B t} \over 2} \bigg ) + 
\gamma \bigg ],
\label{wy3}
\end{equation}
where $\gamma = 0.577215$ is Euler's constant.

As already mentioned the general formula has 
been derived on the basis of impact parameter 
representation of the scattering amplitude
for the first time by Franco \cite{fra1},
which has been further developed in Refs.
\cite{lapi,cahn,kun1}. The given approach
has opened some new important questions concerning
the interpretation of hadronic elastic scattering;
we will explain it to a greater detail in Sect. {\ref{sec2}}.
Sect. {\ref{sec3}} will be then devoted to comparing both
the approaches; especially the limitations contained in
the former one and not yet sufficiently analyzed 
will be discussed. The importance of the latter
approach for analyzing experimental data will
be discussed in Sect. {\ref{sec4}}.

\section{Approaches based on the impact parameter representation}
\label{sec2}
The mentioned approaches using the impact parameter
representation of the scattering amplitudes are
based on the eikonal models. The form suggested by
Glauber \cite{glau} has been used
\begin{equation}
F(s,q^2=-t)= {s\over {4 \pi i}} \int\limits_{\Omega_b}d^2b
e^{i\vec{q}\vec{b}} \bigg[e^{2i\delta(s,b)}-1\bigg]
\label{ei1} 
\end{equation}
where $\Omega_b$ represents the two-dimensional 
Euclidean space of the impact parameter $\vec b$.
A mathematically correct formulation of the impact
parameter theory (respecting fully the existence of 
a finite interval of admissible 
$t$ values at finite energies) was given by
Adachi et al. \cite{adac} and generalized by Islam
\cite{isla,islb} who showed that it is valid at any $s$ and
$t$. In the elastic scattering of two charged nucleons the 
corresponding eikonal can be written (due to additivity 
of the potentials \cite{fra1,islb}) as
\begin{equation}
\delta^{C+N}(s,b)= \delta^{C}(s,b) + \delta^{N}(s,b),
\label{ei2}
\end{equation}
and Eq. (\ref{ei1}) for the total elastic scattering 
amplitude may be rewritten as
\begin{equation}
F^{C+N}(s,t=-q^2) = {s\over{4\pi i}} \int\limits_{\Omega_b} d^2b
e^{i\vec{q} \vec{b}}\bigg[e^{2i(\delta^{C}(s,b)
+\delta^{N}(s,b))}-1\bigg].
\label{ei3}
\end{equation}
Eq. (\ref{ei3}) can be then transformed according  
to Refs. \cite{fra1} and \cite{cahn} 
(see also \cite{kun1}) into the form
\begin{eqnarray}
 F^{C+N}(s,t) \!\! &=& \!\! F^{C}(s,t) + F^{N}(s,t) +
 {s\over{4\pi i}} \int\limits_{\Omega_b} d^2b
e^{i\vec{q} \vec{b} }\bigg[e^{2i\delta^{C}(s,b)}-1\bigg]
\bigg[e^{2i\delta^{N}(s,b)}-1\bigg]
\nonumber\\
&=& \!\! F^{C}(s,t) + F^{N}(s,t) +
{i\over {\pi s}} \int\limits_{\Omega_{q'}} d^2q'
F^{C}(s,{q'^2})F^{N}(s,[\vec{q} - \vec{q'}]^2),
\label{ei4}
\end{eqnarray}
where $F^{C}(s,t)$ and $F^{N}(s,t)$ are Coulomb and elastic 
hadronic amplitudes defined by expression (\ref{ei1}) with
the eikonals $\delta^{C}(s,b)$ and $\delta^{N}(s,b)$. Equation 
(\ref{ei4}) describes simultaneous actions of both
the hadron and Coulomb forces responsible for 
the total elastic scattering. It includes the 
convolution integral of the two amplitudes
defined over kinematically allowed region 
of momentum transfers $\Omega_{q'}$.
It means that the total elastic amplitude
(\ref{ei4}) can be expressed as the sum 
of both the Coulomb and hadronic amplitudes
to which a function depending on both the 
Coulomb and hadronic amplitudes is added.

Franco and Cahn have started from Eq. (\ref{ei4}),
but they have passed to some simplifications; 
the impact parameter formalism used in their 
approaches has been valid at asymptotic energies and
small momentum transfers only. They have tended mainly 
to re-deriving the West and Yennie formula on the basis
of eikonal formalism. 

The general formula ({\ref{ei4}) holds at any $s$ and $t$ 
and may be further reformulated \cite{kun1}.
Using the generalized impact parameter representation
formalism introduced by Adachi and Kotani \cite{adac} 
and Islam \cite{isla,islb} 
it has been possible to derive the expression for 
the total elastic scattering amplitude valid generally
up to the terms linear in $\alpha$.   
It has been possible to write \cite{kun1}
\begin{equation}
F^{C+N}(s,t) = \pm {\alpha s\over t}f_1(t)f_2(t) +
F^{N}(s,t)\Bigg [1\mp i\alpha G(s,t) \Bigg ],  
\label{kl1}
\end{equation}
where
\begin{equation}
G(s,t) = \int\limits_{t_{min}}^0
dt'\Bigg \{ \ln{t'\over t}\bigg[f_1(t')f_2(t')\bigg]'
+ {1\over {2\pi}}\Bigg [{F^{N}(s,t')\over F^{N}(s,t)}-1\Bigg]
I(t,t')\Bigg \},
\label{kl2}
\end{equation}
and
\begin{equation}
I(t,t')=\int\limits_0^{2\pi}d{\Phi^{\prime \prime}}
{f_1(t^{\prime \prime})f_2(t^{\prime \prime})\over t^{\prime \prime}};
\label{kl3}
\end{equation}
here $t^{\prime \prime}=t+t'+2\sqrt{tt'}\cos{\Phi}^{\prime \prime}$.
For the case of nucleon - nucleon scattering $t_{min}=-s + 4 m^2$.
The upper (lower) sign again corresponds to the $pp$
($\bar{\mbox{p}}p$) scattering.

Differential cross section can be then defined as
\begin{equation}
{{d \sigma (s,t)} \over {dt}}= {\pi\over {sp^2}}|F^{C+N}(s,t)|^2.
\label{ds1}
\end{equation}
Making use of the optical theorem one can write for the
total cross section 
\begin{equation}
\sigma_{tot} (s) = {{4 \pi}\over {p \sqrt{s}}} \Im F^{N}(s,t=0).
\label{ot1}
\end{equation}
Instead of the $t$ independent quantities
$B$ and $\rho$, it is now necessary to define
$t$ dependent quantities
\begin{equation}
B(s,t)= {d\over {dt}}\bigg[\ln {d \sigma^{N}\over {dt}}\bigg] =
{2\over |F^{N}(s,t)|}{d\over {dt}}|F^{N}(s,t)|
\label{sl1}
\end{equation}
and
\begin{equation}
\rho (s,t) = {{\Re F^{N}(s,t)} \over {\Im F^{N}(s,t)}}.
\label{ro2}
\end{equation}
Assuming for both the last quantities to be 
$t$ independent corresponds to a fundamental 
limitation of formula (\ref{wy2}) and disables
practically its use for model interpretations.
However, some important limitation has been 
contained already in Eq. (\ref{wy1}).
All problems relating to Eqs. (\ref{wy1})
and (\ref{wy2}) will be analyzed in the 
following section.

\section{Limitations involved in West and Yennie approach}
\label{sec3}

Leaving aside neglection of spins (assumed 
in both the approaches) it has been believed that 
the original West and Yennie integral formula 
for relative phase (\ref{wy1}) does not contain 
practically any limitations concerning the $t$ 
dependence of the hadronic amplitude $F^N(s,t)$; and that
only its simplified form expressed by Eq. (\ref{wy2})
was based on limiting assumptions.
However, it is not true as one important limitation follows
already from requiring for the relative phase (\ref{wy1}) 
to be real; the following relation
\begin{equation}
\rho(s,t)= {{\Re F^N(s,t)} \over {\Im F^N(s,t)}}
= {{\Re F^N(s,t')} \over {\Im F^N(s,t')}} = \rho(s)
\label{wy4}
\end{equation}
between the real and imaginary parts of $F^N(s,t)$ 
should be fulfilled for any values of $t$ and $t'$.
Thus the West and Yennie
integral formula (\ref{wy1}) admits only such $t$ dependence of
the hadronic amplitude $F^N(s,t)$ which leads to constant
value of the ratio $\rho(s)={{\Re F^N(s,t)} \over {\Im F^N(s,t)}}$
{\it{in the whole region of kinematically allowed $t$ values}}.
As the quantity $\rho(s)$ in the case of elastic nucleon
collisions is small the West and Yennie integral formula
(\ref{wy1}) admits only the elastic hadronic amplitude 
with dominant imaginary part in the whole kinematically
allowed region of $t$.

The $t$ dependence of differential cross section 
and diffractive slope are then given according 
to Eqs. (\ref{ds1}) and (\ref{sl1}) fully by the 
imaginary part of hadronic amplitude only, i.e., 
\begin{equation}
{{d \sigma (s,t)} \over {dt}}= {\pi\over {sp^2}}
(1 + \rho(s)^2) (\Im F^{N}(s,t))^2
\label{ds2}
\end{equation}
and
\begin{equation}
B(s,t)= {2\over |\Im F^{N}(s,t)|}{d\over {dt}}|\Im F^{N}(s,t)|.
\label{sl2}
\end{equation}

Further additional assumption is then contained 
in the simplified West and Yennie relative phase 
(\ref{wy3}). While Eq. (\ref{wy1}) does not
exclude for the diffractive slope $B(s,t)$
to be $t$ dependent, the Eq. (\ref{wy2})
can hold only if $B(s,t) = B(s)$, i.e., if the
differential cross section or the modulus 
$|F^{N}(s,t)|$ of the hadronic amplitude 
is purely exponential \cite{kun2}. The
elastic hadronic amplitude is then given
by the second term of Eq. (\ref{wy2}).

We will show now to a greater detail the
approximations being included in derivation 
of all previous formulas and assertions. Let us
denote the integral in 
Eq. (\ref{wy1}) by $A$. Then
\begin{eqnarray}
A &=& \int_{-4p^2}^{0}{dt'\over |t'-t|}
\bigg[1-{F^N(s,t')\over F^N(s,t)}\bigg]   
= \int_{-4p^2}^{0}{dt'\over |t'-t|}
[1- e^{B (t' - t)/2} ] =  
\nonumber \\
&=& \int_{0}^{(4p^2+t)B/2}{dy\over y}
[1- e^{-y} ] +
\int_{0}^{-Bt/2}{dy\over y}
[1- e^{y}]=
\nonumber \\
&=&E_1(B/2(4 p^2+t)) + \ln (B/2(4 p^2 + t)) + \gamma  
\nonumber \\
&-& Ei(-Bt/2) + \ln (-Bt/2) + \gamma,
\label{wy7}
\end{eqnarray}
where  formulas (5.1.39) and(5.1.40) from 
Ref. \cite{abra} have been applied to. The 
exponential integrals $E_1(z)$ and $Ei(x)$ are 
defined, e.g., by Eqs. (5.1.1) and (5.1.2) from 
Ref. \cite{abra}. Introducing further two other 
simplifications (asymptotic energies and small 
$|t|$) which may be applied to (as West and Yennie 
did; see Ref. \cite{west}) the integral 
$A$ equals  
\begin{equation}
A = \ln(Bs/2) + \gamma,
\end{equation}
which gives finally the
simplified West and Yennie formula 
(\ref{wy3}) for the relative phase;
the asymptotic expansions for 
the involved exponential integrals 
(Eqs. (5.1.50) and (5.1.51) from Ref. \cite{abra})
being applied to.

Thus the simplified formula (\ref{wy3}) involves
a series of limiting assumptions:
(i) constant $\rho$; (ii) purely exponential $t$ dependence
of the modulus; (iii) asymptotic energies; (iv) small $|t|$. 
It means that it is in principle inconsistent with 
actual experimental data. 

Some important discrepancies concerning the experimental
characteristics have followed already from $t$ independence
of $\rho$. If $\rho$ is constant at the given energy  
its derivative in the $t$ variable is zero. 
Then it follows 
\begin{equation}
{d \over {dt}} \Re F^N(s,t) \;\;\;\Im F^N(s,t) =
\Re F^N(s,t)\;\;\; {d \over {dt}} \Im F^N(s,t)
\label{wy9}
\end{equation}
for all admissible values of $t$.
On the other hand the existence of diffractive minimum
observed in all elastic nucleon collisions leads to
a condition that the first derivative of 
differential cross section should be zero. 
It holds then
\begin{equation}
\Re F^N(s,t_D) \;\;\;{d \over {dt}} \Re F^N(s,t_D) =
-\Im F^N(s,t_D) \;\;\;{d \over {dt}} \Im F^N(s,t_D);
\label{wy10}
\end{equation}
the corresponding diffractive minimum being at $t_{D}$.
It follows from Eqs. (\ref{wy9}) and (\ref{wy10}) 
that both the real and imaginary parts of $F^N(s,t_D)$
should equal zero, which contradicts the experimental 
data as the differential cross section does not vanish
in the diffractive minimum. The existence of diffractive 
minimum observed in all diffractive hadron collisions
is, therefore, in a clear contradiction to $\rho$ 
being constant.

As to the experimental data it is, of course, also the
assumption concerning the purely exponential
$t$ dependence of the modulus $|F^N(s,t)|$ at
{\it{all kinematically allowed values of momentum transfers}}
that is in contradiction to the present high energy
elastic nucleon scattering data. 
It has been observed experimentally that for
the $pp$ elastic scattering at the ISR energies the 
region of {\it{approximately}} exponential $t$ dependence of
differential cross section is only for $t$ running from
the forward direction to the diffractive minimum. The values of 
${{d \sigma} \over {dt}}$ change here within 8 or 9 orders of 
magnitude. And this region becomes narrower and the range 
of the corresponding magnitudes becomes smaller when the 
collision energy increases, e.g., for $\bar{p}p$ scattering
at the Collider energy 541 GeV the magnitude change is only 
5 orders. And the model predictions for $pp$ scattering at
the LHC energy 14 TeV tell us that this change will be 
approximately only 3 orders of magnitude (see, e.g., Ref. \cite{tdr}).
At higher $|t|$ the secondary maximum appears that clearly
indicates the modulus not to exhibit purely exponential 
behavior in $t$ in the interval of all admissible $t$ values. 

However, in the past (before ISR experiments) nothing
was known about the existence of diffractive minima in
elastic hadron collisions. The differential cross section 
data exhibited only purely exponential $t$ dependence in
the rather narrow regions of studied momentum transfers.
Therefore, the use of 
simplified West and Yennie amplitudes for analysis 
of elastic scattering data at lower energies could be
regarded as justified at that time.

Having been aware of great discrepancies of 
formula (\ref{wy2}) in the region of higher
$|t|$ people started to use two different 
formulas: Eq. (\ref{wy2}) for the region 
of very small $|t|$ and some phenomenological
formula for the whole other interval of $|t|$;
for detail see, e.g., Refs. \cite{baro,donn}.

It has been assumed that the triple of quantities
$\sigma_{tot}, \rho, B$ may characterize an actual
hadronic amplitude in the region around $t \sim 0$.
However, neither this assumption may be regarded as
justified as Eq. (\ref{wy2}) has been derived by
integrating over the whole kinematically allowed
interval of $t$ under the conditions differing 
drastically from reality. Consequently, only the
general formula (without all mentioned simplifications
and limitations) may be efficient in interpreting 
available experimental data.

We should like to repeat that the simplified
West and Yennie amplitude (\ref{wy2}) is used
in an inconsistent way even if it is applied to the 
interference region only. 
This discrepancy cannot be removed even if a more 
general shape is applied to other regions of the
measured differential cross section. It is also this 
application of two different formulas (one for interference
region and a different one for hadronic region - based 
on some phenomenological approach; for detail see, e.g., 
Refs. \cite{baro,donn}) that represents an important deficiency.
All shortages and discrepancies may be removed, however, by using 
one common eikonal formula, only.

Several authors \cite{cahn,sely,kope} tried to improve
formula (\ref{wy2}) by calculating
the next-to-leading or next-to-next-leading order
terms to it which might be important at higher $|t|$.
The mentioned authors wanted to derive
the {\it{analytical}} expressions for the relative
phase between the Coulomb and hadronic elastic amplitudes.
In tending to it they have had to assume the
exponential $t$ dependence of hadronic amplitude
and of the form factors at all kinematically 
allowed $t$ values (in order to perform the 
analytical calculation of the corresponding 
integrals). However, such assumptions 
do not correspond to the actual 
behavior of the experimental data and can
depress the importance of higher corrections
and their influence on the relative phase.
This has been confirmed by recent analysis
of experimental data (see Ref. \cite{petr})
when the used different formulas for the relative
phase gave approximately the same results. 

\section{General formula and different interpretations 
of experimental data}
\label{sec4}

There is not any actual theory of elastic
hadronic nucleon scattering and the shape of 
elastic hadronic amplitude $F^{N}(s,t)$ must be 
derived from experimental data concerning the 
$t$ dependence  of differential cross section.
However, the complex elastic hadronic amplitude
\begin{equation}
F^{N}(s,t) = i |F^{N}(s,t)| e^ {-i \zeta^{N} (s,t)},
\label{nu1}
\end{equation}
is characterized by a pair of the real functions,
i.e., by the modulus $|F^{N}(s,t)|$ and the phase
$\zeta^{N} (s,t)$; it holds
\begin{equation}
\rho(s,t) = {{\Re F^{N}(s,t)} \over {\Im F^{N}(s,t)}} =
\tan {\zeta^{N}(s,t)}.
\label{ro3}
\end{equation}
However, for this pair of functions only
one experimentally determined function 
${{d \sigma } \over {dt}}$ is available.
Thus, the complete form of scattering amplitude
cannot be derived from experimental data and
some other arguments for its definite form
must be looked for.

It is the distribution of processes
in the impact parameter space that 
depends on the function $\zeta^{N}(s,t)$
in a decisive way. This distribution 
$D(s,b), b \ge 0$ is given by the Fourier-Bessel 
transformation
\begin{equation}
h_{el} (s,b) = {1 \over {4 p \sqrt{s}}} 
\int \limits_{t_{min}}^{0} \! dt \; F^{N}(s,t)\; J_{0}(b \sqrt{-t});
\label{el1}
\end{equation}
it holds
\begin{equation}
D(s,b) = |h_{el}(s,b)|^2.
\label{el2}
\end{equation}
Assuming for $\rho(s,t)$ to hold $\rho (s,0) \ll 1$ in Coulomb 
and interference regions and to increase monotony for higher
values of $|t|$ in such a way that $\Im F^{N}(s,t)$
vanishes in the diffractive minimum, e.g., 
$\zeta^N(s,t_{D}) = {{\pi} \over 2}$, the amplitude 
$F^N(s,t)$ may be fully derived from experimental data. 
In such a case the function $D(s,b)$ has a Gaussian
shape with the maximum at $b=0$. And we can speak about the
central behavior of elastic hadronic collisions, which
attributes, e.g., to proton the structure differing 
fundamentally from that required normally for diffractive 
production collisions. 

In such a diffractive case the function $D(s,b)$ 
should have the maximum at $b > 0$ and we can speak
about peripheral behavior. The corresponding
function $D(s,b)$ may be easily obtained 
if the function $\zeta^{N} (s,t)$ having a
small non-zero value at $t=0$ increases quickly
with rising $|t|$. The more detailed description of 
such a case can be found in Refs. \cite{kun3,kun4,kun5,kun6}.
Similar behavior of the phase $\zeta^{N} (s,t)$
was considered by Franco and Yin \cite{fra3}
for nuclear collisions.

The analyses of experimental data 
for $pp$ scattering at the energy of 53 GeV
and for $\bar{p}p$ scattering at energy of 541 GeV
based on different additional assumptions  
may be found in Ref. \cite{kun1}. The following 
conclusion should be done from the given analysis:

\begin{itemize}

\item{the diffractive slope $B(s,t)$ and the
$\rho(s,t)$ quantity are $t$ dependent in the whole
measured $t$ region in the peripheral as well as 
in the central behavior of elastic hadron scattering}

\item{the influence of Coulomb scattering
can be hardly neglected at higher $|t|$ values}

\item{the peripheral feature of elastic nucleon scattering
at high energies seems to be slightly statistically preferred}

\item{the use of the total elastic amplitude (\ref{kl1})
should be strongly supported}

\end{itemize}

In the end of this section we should like to mention yet
one recent attempt of solving the interference problem on
the basis of simplified formula (\ref{wy2});
see Ref. \cite{gaur}. It is possible to write 
in such a case
\begin{equation}
{{\pi}\over {sp^2}}{{d \sigma} \over {dt}} =
 \bigg [ \bigg ( F^C + \Re F^N \bigg )^2 +
 \bigg ( \alpha \Phi F^C + \Im F^N \bigg )^2 \bigg]. 
\label{ni1}
\end{equation}
In the case of $pp$ elastic scattering at high energies
the $\Re F^N$ is small and the first term on the right hand 
side should tend to zero already in interference region; 
this zero value being reached at $t_{min}$. And the
differential cross section at this point should be 
given by the other term. Combining experimental data 
with remaining theoretical values the value of $t_{min}$ 
may be derived. As the remaining part of the equation
has not been fulfilled authors of Ref. \cite{gaur} 
have introduced an approach of bringing
it to zero and claimed that they established
final values of $\sigma_{tot}$, $\rho$ and $B$.
However, it does not represent any actual solution
of the collision pattern as the formula used is 
burdened by large approximations. In such a case 
one should actually write 
\begin{eqnarray}
{{d \sigma} \over {dt}} &=& {{\pi} \over {s p^2}} 
|F^C + F^N (1 -i \alpha G)|^2
\nonumber \\
&=& {{\pi} \over {s p^2}} \bigg [ \bigg ( 
F^C + \Re F^N + \alpha
 (\Re F^N \Im G + \Im F^N \Re G) \bigg )^2 
 \nonumber  \\
 &+&\bigg ( \Im F^N + \alpha 
 (-\Re F^N \Re G + \Im F^N \Im G)\bigg )^2 \bigg ]  
\label{ni2}
\end{eqnarray}
or
\begin{eqnarray}
{{sp^2} \over {\pi}} {{d \sigma} \over {dt}} &-&
\bigg ( \Im F^N + \alpha 
 (-\Re F^N \Re G + \Im F^N \Im G)\bigg )^2   =
 \nonumber \\
&=&  \bigg ( 
F^C + \Re F^N + \alpha
 (\Re F^N \Im G + \Im F^N \Re G) \bigg )^2 \equiv \Delta_R^2 (t). 
\label{ni3}
\end{eqnarray}
The expression of the right hand side of Eq. (\ref{ni3})
exhibits some deep minima (even if never equal zero).
The behaviors corresponding to peripheral and central 
cases of the $pp$ elastic scattering at energy of 53 GeV
are shown in Fig. 1; results being based on the analysis 
of data given in Ref. \cite{kun1}. There are 
pronounced minima, their positions being 
different for central and peripheral behaviors. The full line
corresponds to the case of peripheral behavior while the other
curves correspond to the central behavior of elastic scattering.
The dashed line corresponds to the central phase used in
Ref. \cite{kun1} while the dotted line corresponds to the
central behavior given by a slightly modified phase of hadronic
amplitude, i.e., $\zeta^N(s,t) = \arctan ( \zeta_1 -  
{{\zeta_0}\over {1 - |{t \over {t_{di\!f\!f}}}|}})$.
The two cases with central behavior exhibit different
numbers of minima.

There is, however, yet greater difference between these two
cases. In the former case the real part $\Re F^N$ does not
passes through zero in the whole kinematically allowed
interval of $t$, while one passage through zero exists
in the latter case. And it may be interesting
that these two solutions exhibit different relations to 
theorems derived by Martin \cite{mart}. The first solution
(dashed line) corresponds to the case where the total 
cross section should remain finite with increasing energy; 
the other solution (dotted line) admits then the rise of the
total cross section to infinity.

However, let us go back to Fig. 1.
The first minimum for peripheral behavior and those 
for the central solution lies in the neighborhood of
$|t| \sim 0.02$ GeV$^2$ as shown in Fig. 2. They correspond
to the condition $\Re F^N = - F^C$, while all other minima 
in Fig. 1 relate to $\Re f^N = 0$ (changes of sign).

\section{Conclusion}
\label{sec5}

In the conclusion let us return once more to the general formula
(\ref{kl1}) where the expression in the last bracket
may be regarded as the first term in the Taylor series
expansion of the exponential $e^{-i \alpha G}$; then
one can write within the same precision
\begin{equation}
F^{N+C}(s,t) = F^{C}(s,t) + F^{N}(s,t) e^{-i \alpha G(s,t)},
\label{to2}
\end{equation}
the form being practically identical with original 
formula of West and Yennie.
However, the $G(s,t)$ (being complex) cannot
be interpreted as a mere phase. The reality required for
$G(s,t)$ would be equivalent to the condition that the 
quantity $\rho(s,t)$ is constant and vice versa.

Thus, the approach of West and Yennie has been burdened
by a significant simplification from the beginning, 
tending immediately to some unjustified conclusions.
A definite answer may be obtained only if the general
formulas (\ref{kl1}) or (\ref{to2})
are used for experimental data interpretation.

Figure captions:

Fig. 1: Representation of the $t$ dependence of 
$\Delta_R^2(t)$ (i.e., of the right hand
side of Eq. (\ref{ni3})) for $pp$ collisions at 53 GeV; 
(i) for peripheral behavior - full line;
(ii) for central behavior:
dashed line for the phase used in Ref. \cite{kun1},
dotted line for the phase mentioned in the text.

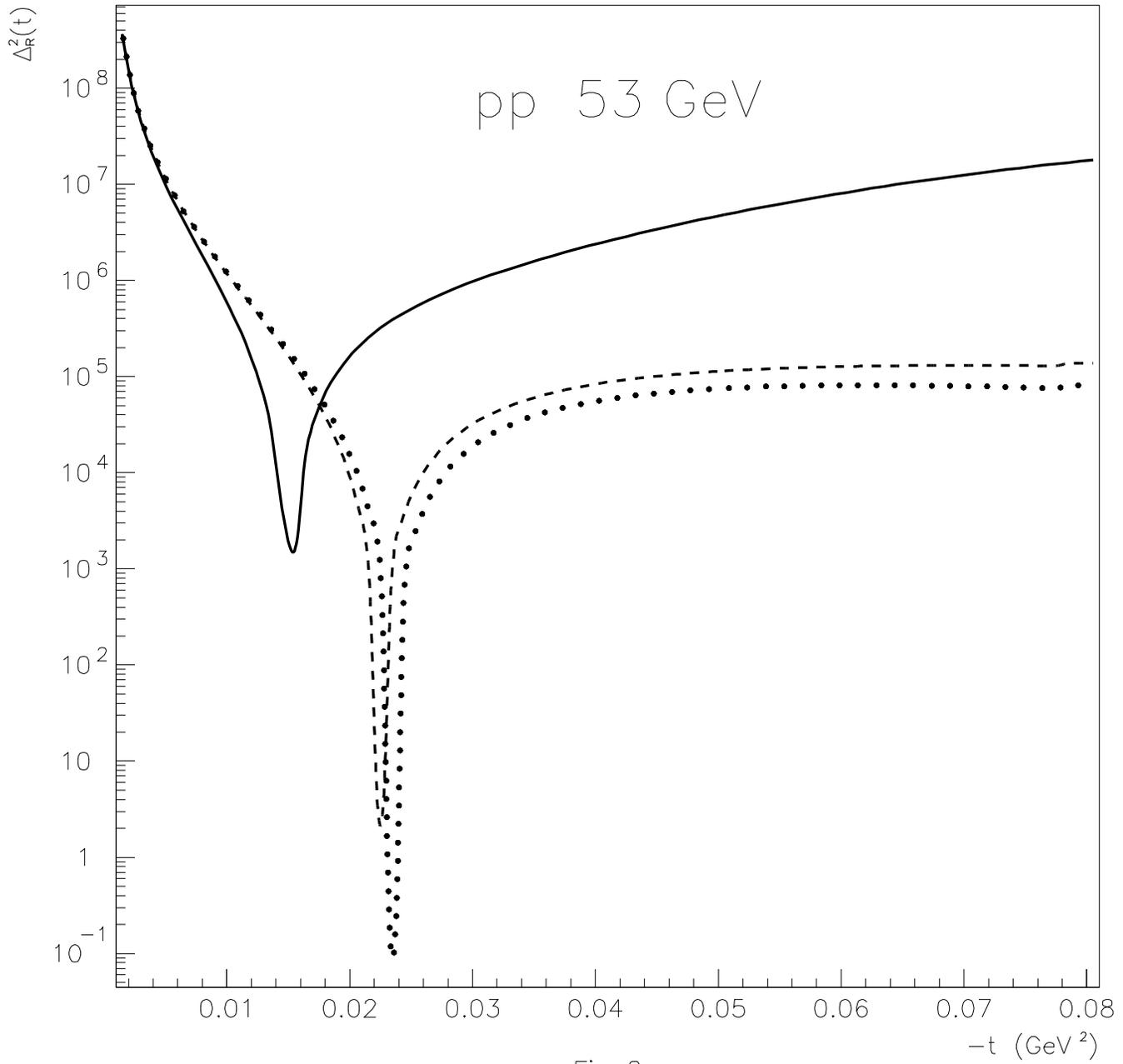
Fig. 2: Representation of the $t$ dependence of
Eq. (\ref{ni3}) $\Delta_R^2(t)$ for $pp$ collisions at 53 GeV. 
Zoomed Fig. 1 for small values of $|t|$.

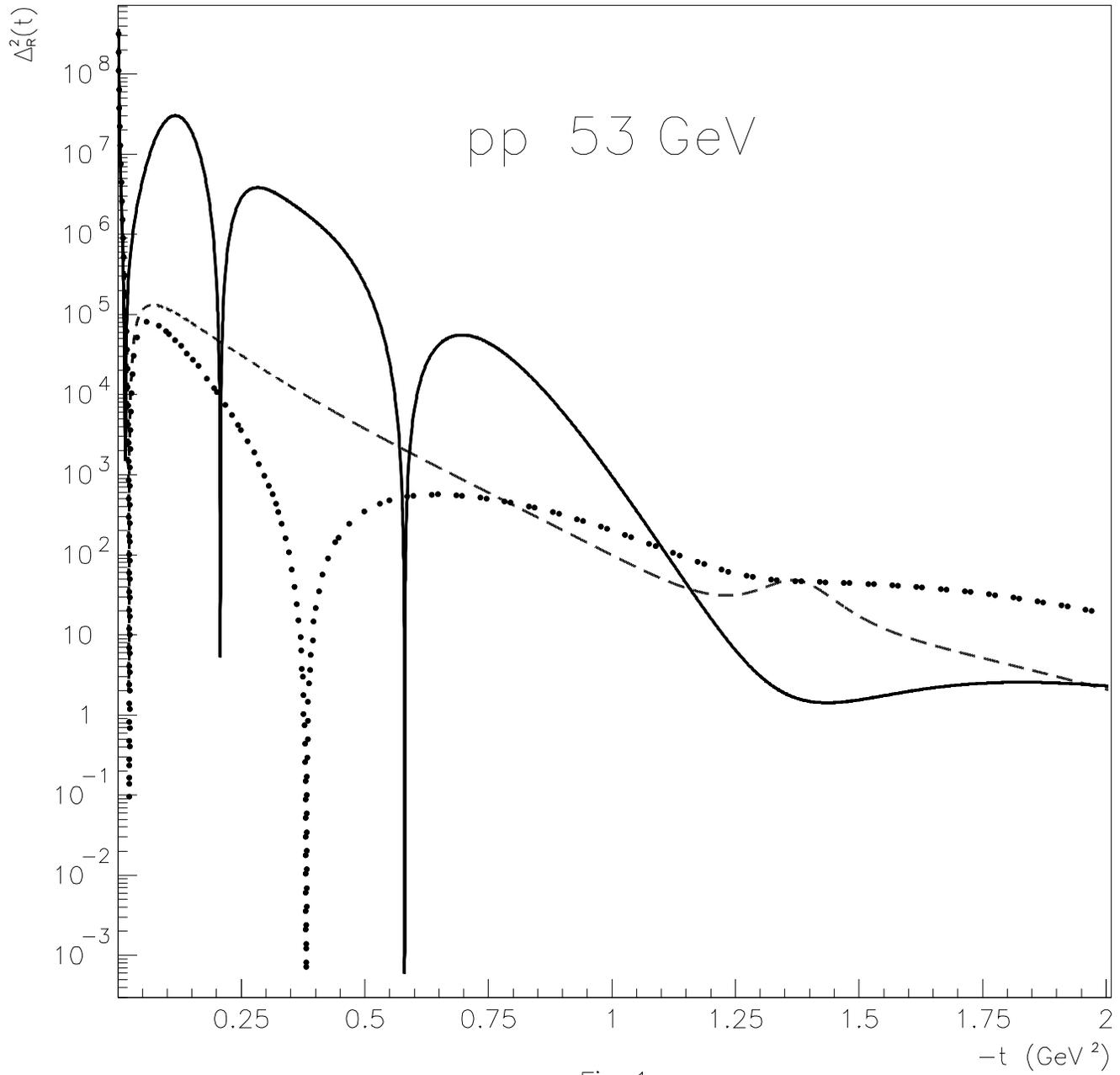
\end{document}